\def\year{2015}
\documentclass[letterpaper]{article}
\usepackage{aaai}
\usepackage{times}
\usepackage{helvet}
\usepackage{courier}
\usepackage{graphicx}
\usepackage{subfigure}
\usepackage{float}

\usepackage{natbib}

\begin{document}
%
\title{Structural Patterns of the Occupy Movement on Facebook}
\author{Michela Del Vicario \\ IMT Lucca, Italy \And
Qian Zhang \\ Northeastern University, Boston, USA \And
Alessandro Bessi \\IUSS Pavia, Italy \And
Fabiana Zollo \\ IMT Lucca, Italy \AND
Antonio Scala \\ ISC CNR,Rome, Italy \And
Guido Caldarelli \\ IMT Lucca, Italy \And
Walter Quattrociocchi \\ IMT Lucca, Italy \\ \textbf{walter.quattrociocchi@imtlucca.it}
}

\maketitle

\begin{abstract}
In this work we study a peculiar example of social organization on Facebook: the Occupy Movement -- i.e., an international protest movement against social and economic inequality organized online at a city level.
We consider 179 US Facebook public pages during the time period between September 2011 and February 2013. 
The dataset includes 618K active users and 753K posts that received about 5.2M likes and 1.1M comments. By labeling user according to their interaction patterns on pages -- e.g., a user is considered to be polarized if she has at least the 95\% of her likes on a specific page --  we find that activities are not locally coordinated by geographically close pages, but are driven by pages linked to major US cities that act as hubs within the various groups. 
Such a pattern is verified even by extracting the backbone structure -- i.e., filtering statistically relevant weight heterogeneities -- for both the pages-reshares and the pages-common users networks.
\end{abstract}


\section{Introduction}\label{intro}

Social media have proved to foster aggregation of people around shared interests such as political ideas, narratives and worldviews \cite{mocanu2014collective,bessi2014science,anagnostopoulos2014viral,friggeri2014rumor}.
An interesting scenario to be explored is the case of online political movements that coordinate and interact through social media \cite{mccaughey2003cyberactivism}. 
To which extent microblogging platforms such as Twitter or Facebook played a crucial role in protest dynamics has been a matter of debate in ~\cite{Garrett, bennett2003communicating, Myers}. 
Occupy Wall Street, an international movement against social and economic inequality, is a peculiar example of the use of Information and Communication Technologies (ICTs). There, the diffusion of ideas, the recruitment of people, as well as the promotion of the protest, are ascribed to online social media~\cite{Chomsky, Bryne}. 
Particular emphasis is given to the simplification gained in the communication and coordination paradigm of protest's activities. The reason why Twitter and Facebook appeared to be particularly suited in supporting the diffusion of political ideas and socio-economic objectives of the movement resides in their structure.
Many works analyzed the topology and dynamics on the Occupy Movement's network \cite{Conover2013-twitter,flammini_digital,gargiulo2014topology,thorson2013youtube} on Twitter. 

However,  users tend to be more active on Facebook. Indeed, the most of online activity during the Arab Spring has been done on Facebook (the total accesses to Twitter were just the 1\% of the entire population) \cite{Tufekci2012}. 
The {\em collective framing} -- the processes that, out of the essential features of the movement's purpose and struggle, establish its narratives, language, and imagery \cite{Garrett,benford2000framing} -- better deal with the Facebook interaction paradigm.
On Facebook users can post information (videos, pictures, etc.) without particular limits and information can be liked (positive feedback), shared, or commented by other users.
In \cite{Neal2011,gaby2012occupy} the diffusion of the Occupy movement on Facebook has been explored. Findings indicate that major uses for Facebook within the movement include the recruitment of people and resources to local occupations, information sharing and story telling, and across-group exchanges.

Taking into account all the posts from $179$ public pages of the Occupy Movement inside the US (US Occupy pages for short) during the period September $2011$-February $2013$, we analyze users' activity on pages and posts. We find high levels of correlation for the number of users, posts, likes, comments and reshares\footnote{The two terms \textit{shares} and  \textit{reshares} are interchangeable throughout the manuscript.} (which are all power law distributed) on each page, and different activity patterns for posts with different post types or shared by users from different categories. Then, we divide users in two categories, {\em habitual} and {\em occasional}, depending on their total number of likes. Furthermore, habitual users having at least $95\%$ of their liking activity on one particular page are defined as {\itshape polarized} in that page. 
Finally, we extract the multi-scale backbone structures for two networks, the {\itshape pages-reshares} network and the {\itshape pages-common users} network. We want to analyze geographical patterns in the information diffusion and polarized users' external activity -- i.e., the activity on all pages but the one where the user is polarized.
Our findings show that the most of the activity, in terms of likes on each page, is made by occasional and not polarized users, and moreover only $21\%$ of polarized users commented at least once. 
However, polarized users tend to increase their probability of commenting multiple times, with respect to general users.
Our analysis points out that, although the majority of posts is shared by common users, they attract little attention i.e., about $6\%$ of total likes. On the contrary, posts shared by page administrators, which represent only $27\%$ of the total, get all the remaining likes. 
These evidences emphasize the central role of the pages in promoting the debate and driving its direction, and at the same time reshape the idea of online debate participation.
We find that online activities are not locally coordinated by geographically close pages, indeed pages linked to major US cities -- e.g., New York, Los Angeles, Chicago, Boston, San Francisco -- drive the diffusion of contents online and serve as coordination points for all other pages, that perform a minor activity in the system. 



\section{Related Work}\label{related_works}
Sociologists and political scientists explored the environment of Internet based social movements focusing on a set of common themes \cite{Garrett,bennett2003communicating,Myers,vanLaer,wray,bennett2008,gaffney2010iranelection,mccarthy,benford2000framing,myers2000media}. In the general frame of communicational and organizational issues, several relevant topics emerged, such as the aforementioned collective framing \cite{Garrett,benford2000framing} and the {\em resources mobilization}, which refers to all those processes exploited by social movements in order to arrange financial, material, and human resources required to sustain their activities in an efficient way \cite{mccarthy}.
The Internet is commonly perceived as a flexible and relatively fluid media, in terms of reorganization. However, at the same time its flexibility may reduce the coherence of the movement's ideological definition. In \cite{myers2000media} three communication eras have been identified, each one dominated by a different communication medium -- i.e., print media and word of mouth, television and the Internet. It is argued that each of those communication technologies implies important differences in the network structures under which potential protestors may be influenced to act and, hence, in the resulting protest waves.
In the particular case of Occupy Wall Street, many works investigated the role of Twitter in organizing and promoting the protest \cite{Conover2013-twitter,flammini_digital,gargiulo2014topology,thorson2013youtube}. The relationship between the geospatial dimensions of social movements communication networks and the organizational pressures facing such movements, was extensively investigated in \cite{Conover2013-twitter}. High levels of geospatial concentration in the attention allocation issue and a differentiation of contents from the national to the international level were found. The evolution of the communication activity and of the topics under discussion was analyzed in \cite{flammini_digital,gargiulo2014topology}.
A study of the evolution of the Occupy Movement on Facebook was presented in \cite{Neal2011}.
Twitter, in addition to Occupy, has also played a prominent role in several other social movements such as the Egyptian revolutionary protests of 2011~\cite{Lotan11,choudhary,gonzalez,Hassanpour2011} and the Arab Spring in general \cite{howard2011opening,khondker2011role}.
The most investigated aspects concern information flows and relationships between news media and information sources \cite{Lotan11,howard2011opening}, analysis of tweets' contents \cite{choudhary}, impact of media disruption on the dispersion of the protest \cite{Hassanpour2011}, presence of social influence and complex contagion in the recruitment patterns \cite{gonzalez}. The role of social ties in movement recruitment were extensively analysed also offline \cite{mcadam1993specifying}. In \cite{Tufekci2012} evidences regarding the prominent role of Facebook with respect to Twitter are reported. A set of $1K$ interviews of people who took part in the protest was analyzed, with the interesting finding that, for about $50\%$ of the interviews, the news of the first demonstrations came from Facebook, face-to-face interaction and telephone. Moreover about $50\%$ of the interviewed declared to have access to Facebook, while it is known that less than $1\%$ had access to Twitter.


\section{Data Description}\label{dataset}

We collected all posts from $179$ Facebook public US pages about the Occupy Movement during the time span September $2011$-February $2013$. Data have been collected using the Facebook Graph API~\cite{API} on public accessible pages, for each of which we have a geographical reference located in the US.
A total of $618K$ users is active, in terms of liking activity, on a set of $753K$ posts\footnote{ For each post we have the following data: post ID, sharing user ID, sharing time, post type -- e.g., photo, status, video or link -- number of received likes and comments, sharing page ID and object ID, that is a unique reference to the post in case of resharing.}. The total number of likes and comments on the downloaded posts are about $5.2M$ and $1.1M$, respectively.
 The dataset represents a complete screenshot of the Occupy Movement in the period immediately following the outbreak of the protest on September 17th, 2011 in the Zuccotti Park of New York. The dataset covers all the posts until the end of February 2013, at the time when all the major protests were no more active. 
 After the Zuccotti occupation, in fact, an October full of similar occupational events followed, leading to an international protest movement that extended itself until the end of 2012, when the movement was principally an online collective protest.

\section{Preliminaries and Definitions}
In this section we introduce the basic notions and definitions that will be used throughout the paper. 
We refer to the set of US Occupy pages and their respective {\itshape active users}  as the {\itshape Occupy System or Network}, where a user is said to be active if she liked at least one post belonging to the Occupy pages.

\paragraph{Statistical Tools.}
We use several figures showing PDFs, CDFs, and CCDFs of some metrics related to activity and consumption patterns. We remind that the probability density function (PDF) of a real--valued random variable is a function $f_{X}$ that describes the probability of the random variable falling within a given range of values, so that $$\mathbf{Pr}[a \leq X \leq b] = \int_{a}^{b}f_{X}(x)dx.$$ The cumulative distribution function (CDF) of a real--valued random variable $X$ is defined as $$F_{X}(x) = \mathbf{Pr}(X \leq x) = \int_{-\infty}^{x}f_{X}(u)du.$$ Similarly, the complementary cumulative distribution function (CCDF) is defined as one minus the CDF, so that $$C_{X}(x) = 1 - F_{X}(x) = \mathbf{Pr}(X > x) = \int_{x}^{\infty}f_{X}(u)du.$$

\paragraph{Classification of users activity.}
We assume the following definitions for the users categories in the Occupy System. 
A first distinction is between {\em habitual} and {\em occasional} users: habitual users are those who liked at least $5$ posts within the Occupy System, while occasional users are those who liked less than $5$ posts. 
Being commonly perceived as positive feedbacks, likes are used as a determinant to identify the membership in one page.
{\em Polarized} users have been identified by a thresholding technique, using again the like as a discriminant. Habitual users having $95\%$ or more of their liking activity on a particular page are said to be polarized on that page.
Notice that a polarized user in the occupy scenario is a user having the most of her liking activity on a specific geolocated community.

\paragraph{Bipartite networks and backbone filter.}
A bipartite graph is a triple $\mathcal{G}=(A,B,E)$ where $A=\left\{ a_{i}\,|\,i=1\dots n_{A}\right\} $ and $B=\left\{ b_{j}\,|\,j=1\dots n_{B}\right\} $ are two disjoint sets of vertices, and $E\subseteq A\times B$ is the set of edges -- i.e. edges exist only between vertices of the two different sets $A$ and $B$. The bipartite graph $\mathcal{G}$ is described by the matrix $M$
defined as
\[
M_{ij}=\left\{ \begin{array}{cc}
1 & if\, an\, edge\, exists\, between\, a_{i}\, and\, b_{j}\\
0 & otherwise
\end{array}\right.
\]

The co-occurrence matrices $C^{A}=MM^{T}$ and $C^{B}=M^{T}M$ count, respectively, the number of common neighbors between two vertices of $A$ or $B$. 
$C^{A}$ is the weighted adjacency matrix of the co-occurrence graph $\mathcal{C}^{A}$ with vertices on $A$. Each non-zero element of $C^{A}$  corresponds to an edge among vertices $a_{i}$ and $a_{j}$ with weight $P_{ij}^{A}$. The co-occurrence graph $\mathcal{C}^{B}$ is built in the same way from the co-occurrence matrix $C^{B}$.

Let $A$ be the set of the $179$ Occupy US pages, $B_1$ the set of all posts equivalence classes' representatives (representatives posts for short)\footnote{We can consider the equivalence relation of having the same object ID; two posts are equivalent, and hence belong to the same equivalence class, if they have the same object ID.}, and $B_2$ the set of all polarized users active on $A$; the {\itshape pages-posts} bipartite network 
is then defined as the triple $\mathcal{G}_1=(A,B_1,E_1)$, where an edge $e_{ij}^1\in E_1$ exists if representative post $b_j^1$ is shared on page $a_i^1$, while the {\itshape pages-polarized users} bipartite network 
is defined as the triple $\mathcal{G}_2=(A,B_2,E_2)$, where an edge $e_{ij}^2\in E_2$ exists if polarized user $b_j^2$ is active on page $a_i^2$.
For our analysis we used two networks derived as co-occurrence networks of  the pages-posts and the pages-polarized users bipartite networks. Considering the co-occurrence matrices $C^{A}_1$ and $C^{A}_2$ we get two co-occurrence networks on the vertex set $A$: the {\itshape pages-reshares} ($\mathcal{C}^{A}_1$) and the {\itshape pages-common users} ($\mathcal{C}^{A}_2$) networks. In $\mathcal{C}^{A}_1$ an edge between two pages exists if at least one representative post is shared on both pages, while in $\mathcal{C}^{A}_1$ there is a link between two pages if they share at least one user who is polarized in either page.

In the last section we show results about the {\itshape Backbone Extraction} presented in ~\cite{Backbone}.  Such a method apply a thresholding filter based on the local identification of the statistically
relevant weight heterogeneities. This kind of approach is able to filter out the backbone of dominant connections in weighted networks with strong disorder, preserving the structural properties and hierarchies at all scales. The Backbone structure has been extracted for the two aforementioned co-occurrence real networks.
The discrimination of the right trade-off between the level of network reduction and the amount
of relevant information preserved in the new representation involve additional issues. In many cases, the probability distribution $P(x)$ that any given link is carrying a weight $x$ is broadly distributed, spanning several orders of magnitude. Such a problem is addressed by using the aforementioned method presented in \cite{Backbone}.


\section{Results and Discussion}
\subsection{Activity on Pages}
\begin{figure}
\centering
\includegraphics[scale=0.40]{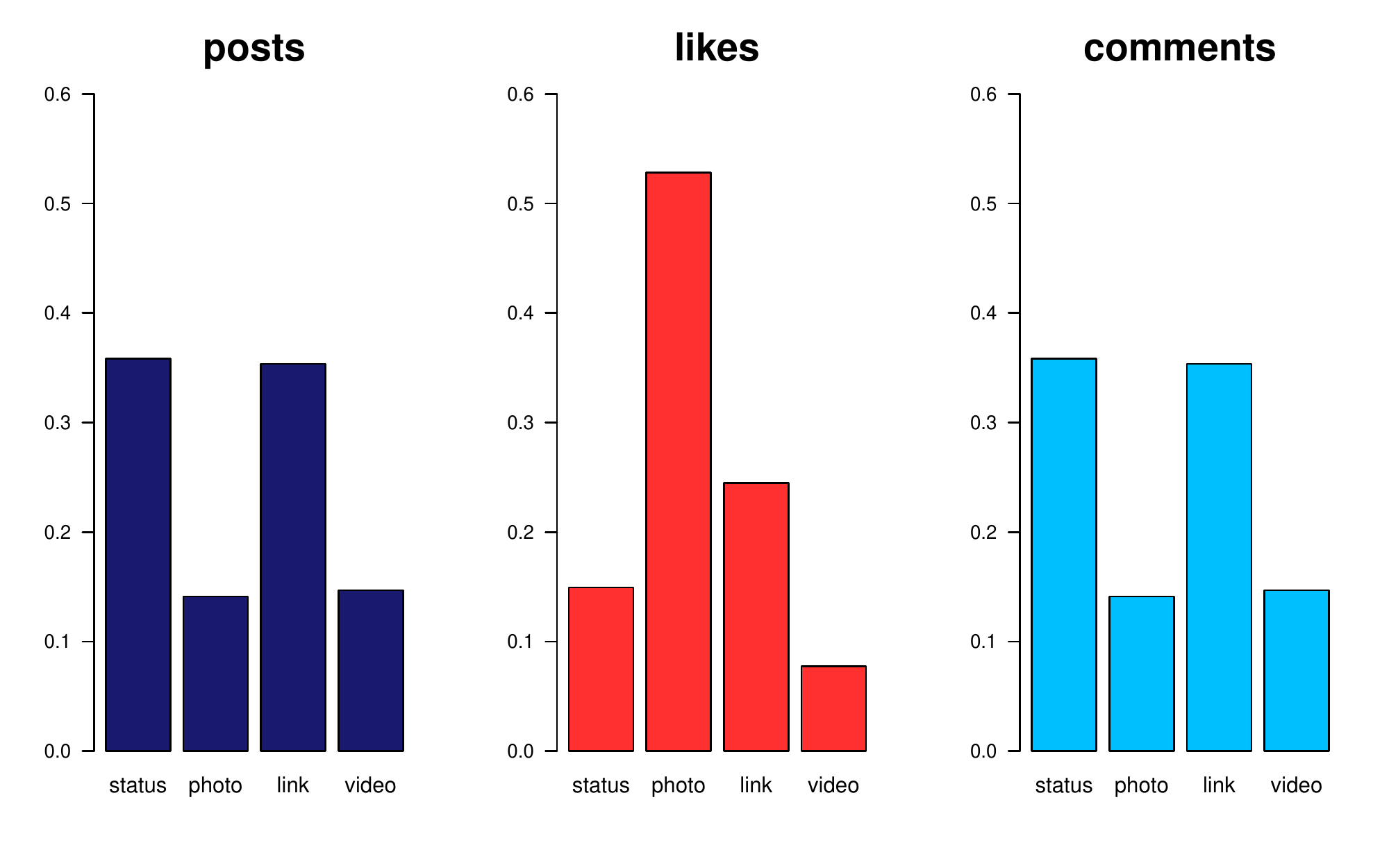}
\caption{\textbf{Fractions of number of posts and activity on posts} (likes and comments) divided by post types. The levels of commenting activities are proportional to those of the number of posts for each different post type. Photos get most part of the likes, although they represent less than a fifth of the total number of posts. For other post types the fraction of likes is one half of the number of posts.}\label{histogram_post_type}
\end{figure}

\begin{figure}
\centering
\includegraphics[scale=0.40]{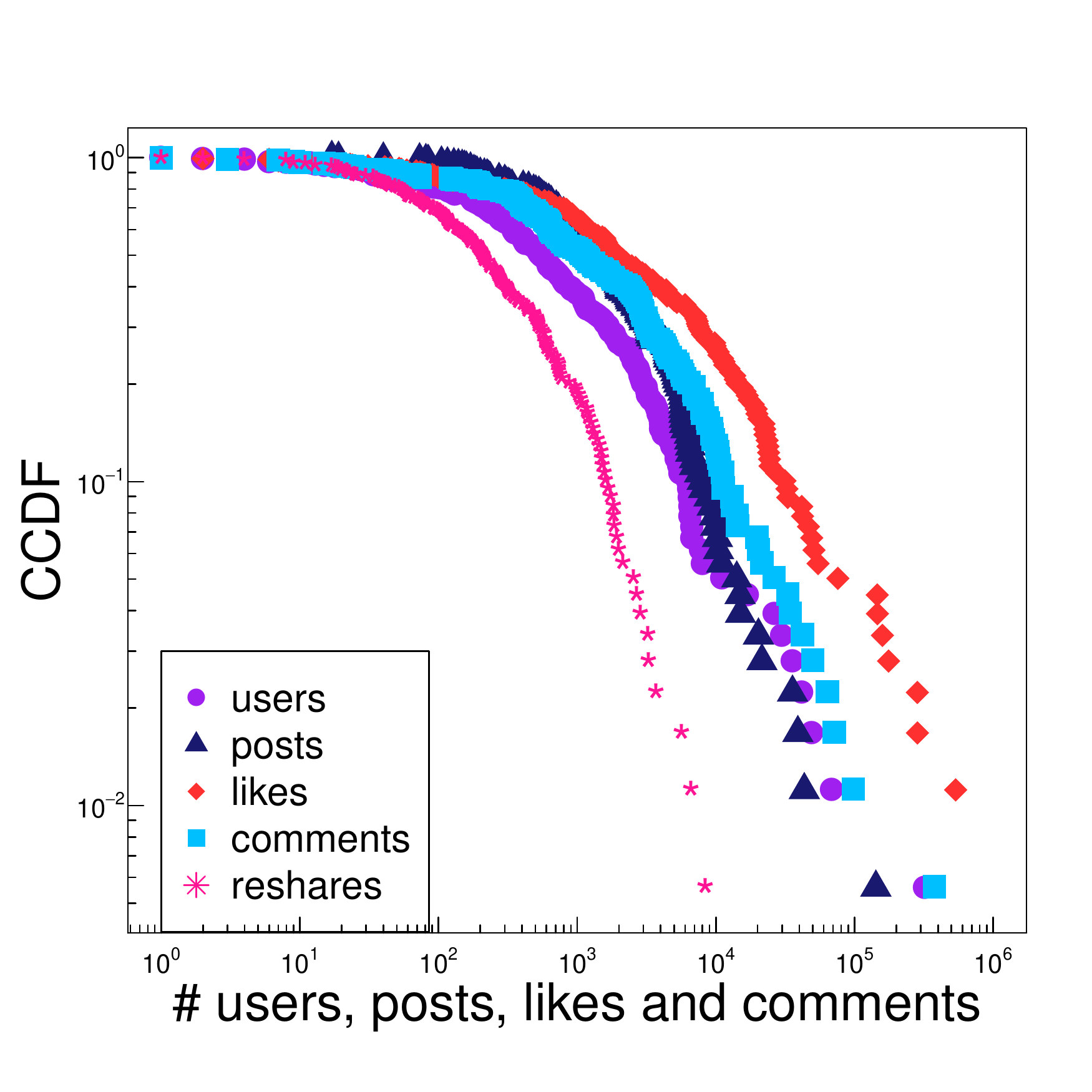}
\caption{\textbf{Empirical CCDF for number of users, posts, likes, comments, and shares on pages.} Taking all the users (purple), posts (light blue), likes (red), comments (sky blue), and shares (pink) on each page we plotted the CCDF of the five quantities and we got a power law distribution for all the measures, with similar shapes and reasonable scale differences.}\label{activity_pages}
\end{figure}

\begin{figure*}
\centering
  \includegraphics[width=0.99\textwidth]{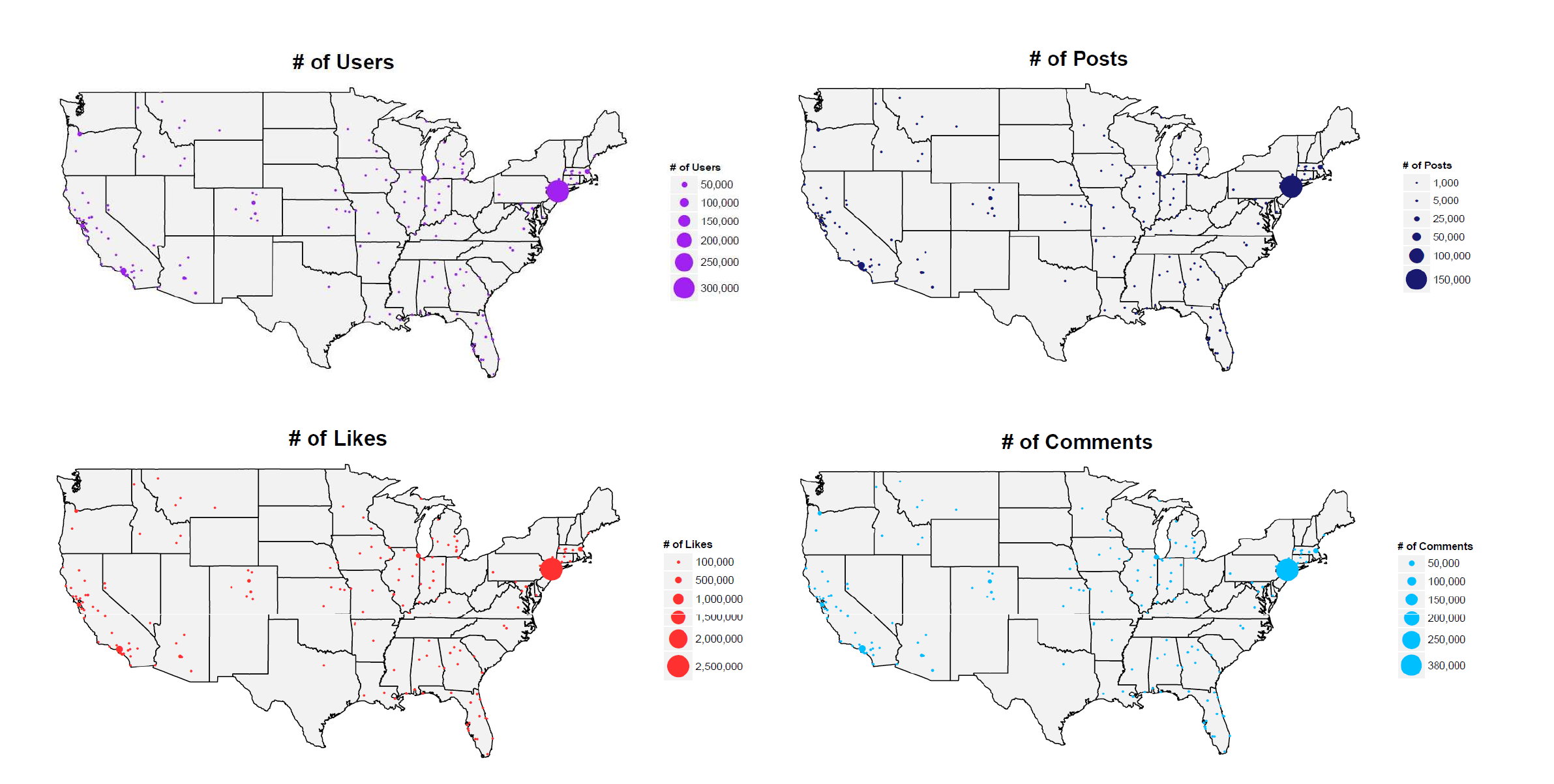}
 \caption{\textbf{Geospatial distributions of activities.} The number of users (purple), posts (light blue), likes (red), and comments (sky blue) is plotted in the form of dots on the US map. The bigger the dot the higher the number of elements in the corresponding category (on the page corresponding to the geographical spot on the map). We can see how the size of the dots, across the different measures, is comparable for all the geographical points. This is a first empirical evidence of the high correlation among the four considered measures.}
\label{maps}
\end{figure*}

Our analysis targets information consumption patterns of the US Occupy Movement. 
The focus is on the relationship between geographic distance and the spreading of information across pages. In addition we characterize the role of polarized users promoting and bridging inter-page connections.
As a first step we look at the users'  activity on the different city related pages. 
Figure \ref{histogram_post_type} reports the fractions of number of posts, likes, and comments for the four kind of posts: status, photo, link, and video. 
The commenting activity is correlated to the number of posts, while more than a half of the likes is concentrated on photos. Statuses, that account for almost two fifths of the total number of posts, get less than a fifth of the total number of likes. 
Such a disproportion in the liking activity might be due to the users limited attention \cite{dukas2001limited,weng2012competition,bessi2014economy} that preferentially tend to select easy to handle post types. 
Figure \ref{activity_pages} shows the empirical complementary cumulative distribution function (CCDF) of the number of users, posts, likes, comments, and reshares for all pages. 
All the five distributions are heavy-tailed. 
To better visualize the geographical distribution of the Occupy Movement, activity patterns are shown on the US map (Figure \ref{maps}). Notice that there exists a high correlation between all pairs of measures -- i.e., users, posts, likes, comments, and shares.

\begin{figure*}
 \centering
\includegraphics[width=0.7\textwidth]{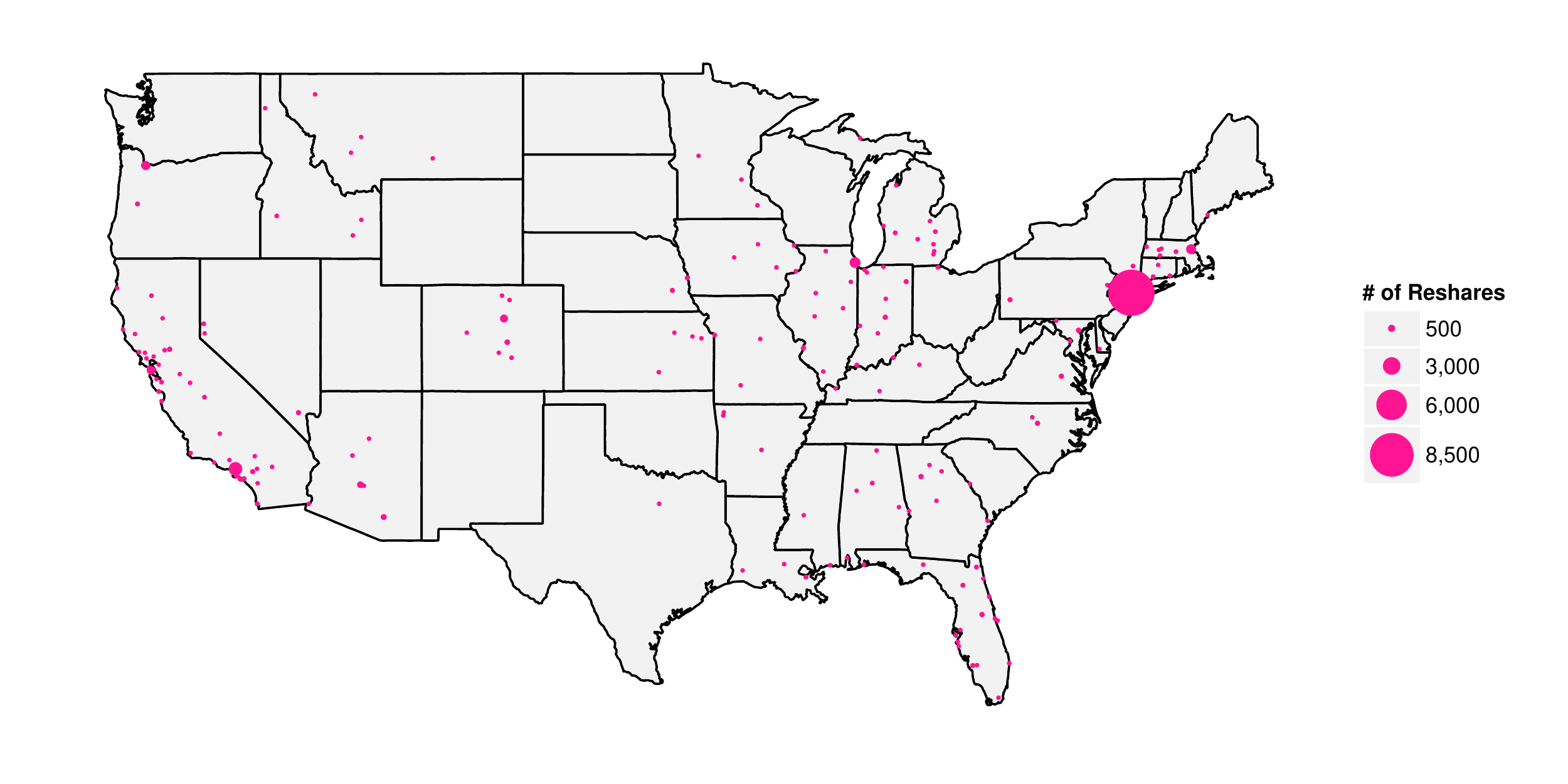}
\caption{\textbf{Resharing patterns.} Results similar to the ones in Figure \ref{maps} hold for the resharing activity. We computed the number of reshares on a subset representing about the $14\%$ of all posts.}
\label{reshares}
\end{figure*}

In Figure \ref{reshares} similar results to those depicted in Figure \ref{maps} hold for resharing patterns. 
\footnote{In order to identify a reshare we need a reference to the previous posts through an Object ID. This is a fundamental information for the reconstruction of resharing patterns, although we had no information about the original posts and no references to the directly reshared posts. }
Pearson correlation coefficient shows high level of correlation between all considered measures -- i.e., number of users, posts, likes, comments and shares -- on the different pages, as shown in Table \ref{correlation_tab}.
Summarizing, the activity of the users on the Occupy pages is comparable across pages; number of users, posts, likes, comments and shares for each page are highly correlated. 

\begin{table}[ht]
\centering
\begin{tabular}{l| c c c c c}
\hline
 & Users & Posts & Likes & Comments & Shares \\ [0.5ex] 
\hline
Users & 1 & 0.974 & 0.997 & 0.991 & 0.739 \\ 
Posts & 0.974 & 1 & 0.963 & 0.987 & 0.840 \\
Likes & 0.997 & 0.963 & 1 & 0.987 & 0.714 \\
Comments & 0.991 & 0.987 & 0.987 & 1 & 0.769 \\
Shares & 0.739 & 0.840 & 0.714  & 0.769 & 1 \\ [1ex]
\hline
\end{tabular}
\caption{\textbf{Correlations between the number of users, posts, likes, comments, and shares.} Pearson correlation for number of users, posts, likes, comments, and shares is high for all different combinations. There are high correlation values among the five considered measures.}\label{correlation_tab} 
\end{table}
	
\subsection{Activity on Posts}
\begin{figure}
\centering
\includegraphics[scale=0.40]{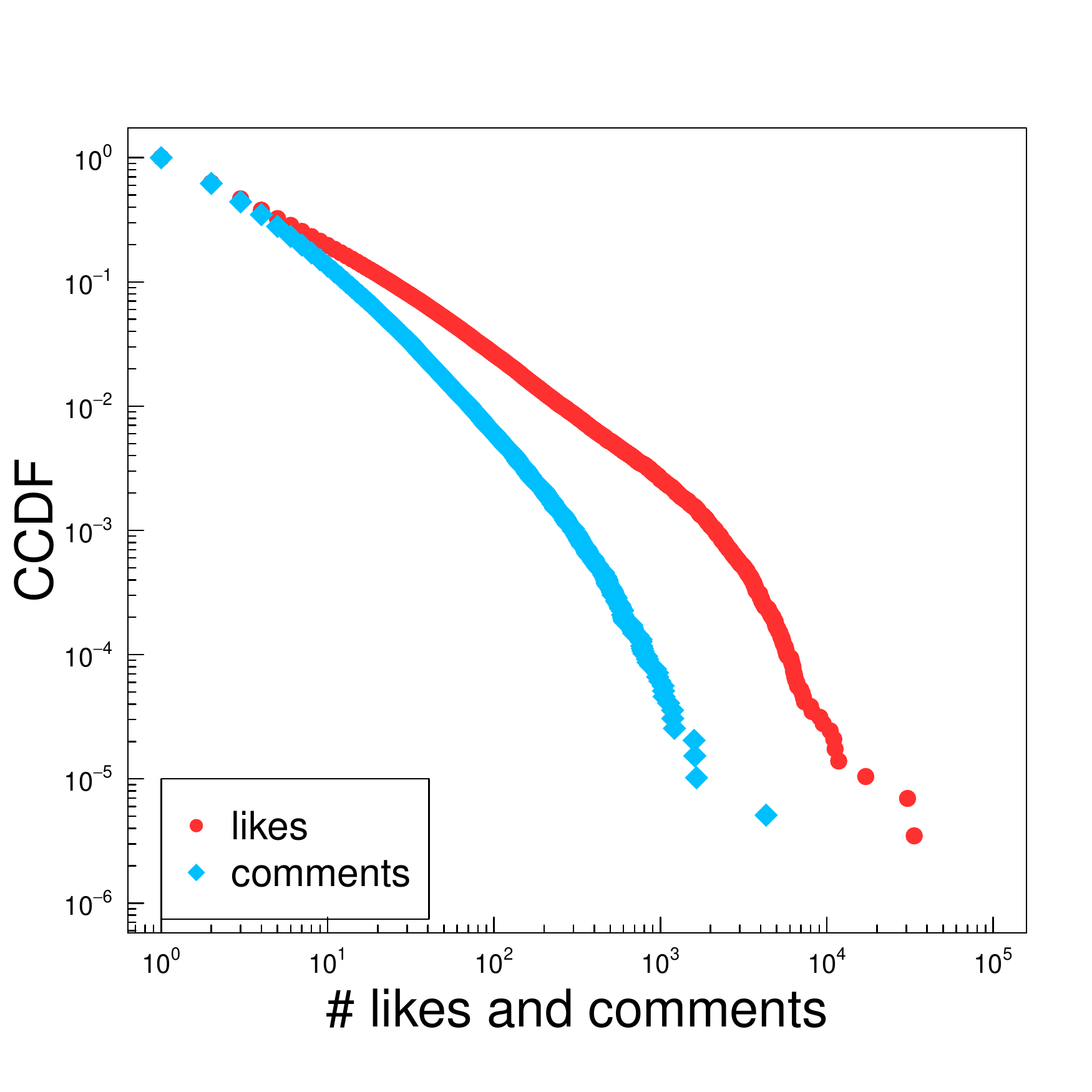}
\caption{\textbf{Empirical CCDF for likes and comments on posts.} The CCDF of the number of likes (red) and comments (sky blue) on the totality of posts, shows power law distributions for the two measures, with a significant difference in the scale.}\label{post_activity1}
\end{figure}

The CCDF of the number of likes and comments for each post shows power law distributions (Figure \ref{post_activity1}). 

\begin{figure}
\centering
\includegraphics[scale=0.40]{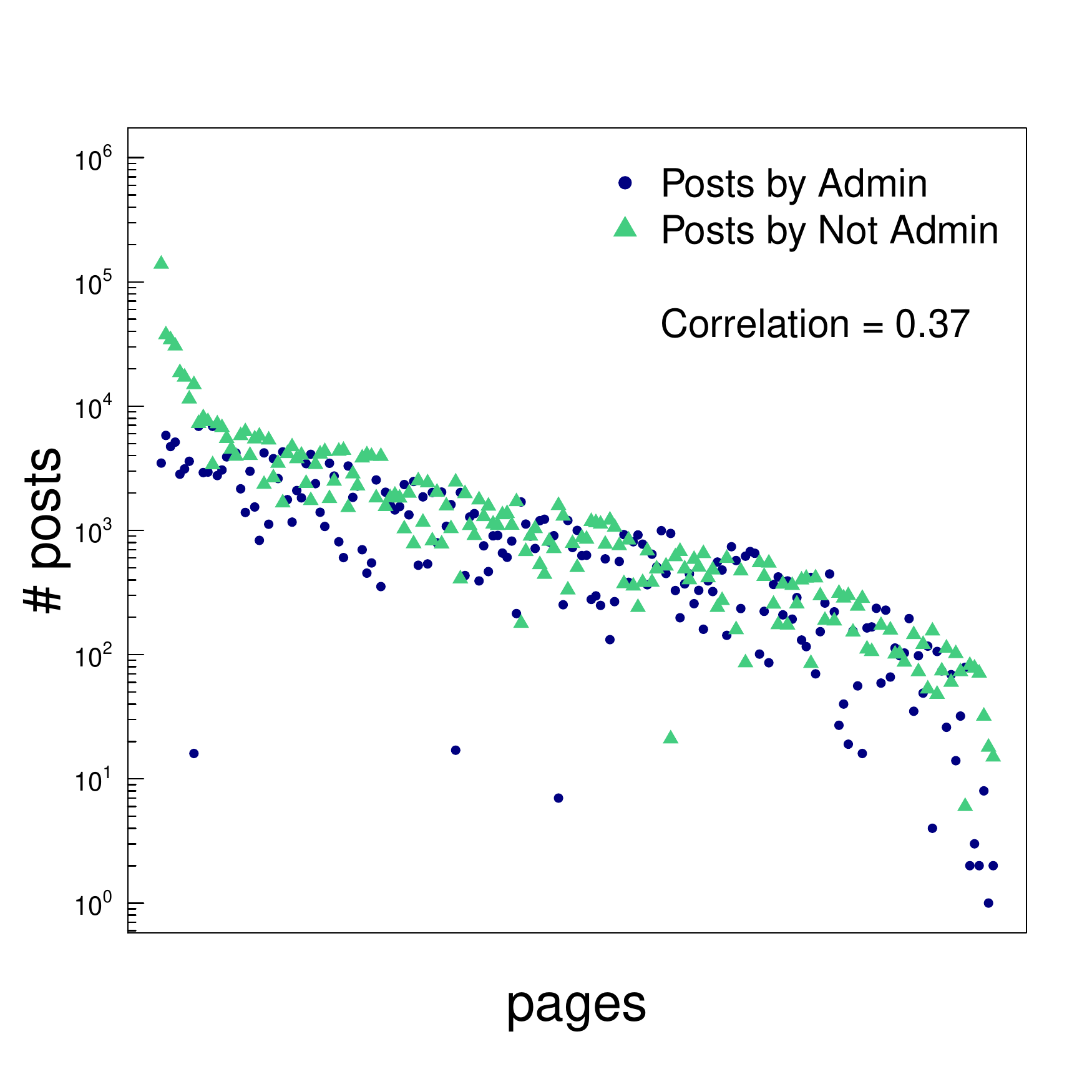}
\caption{\textbf{Posts by admin and not admin.} On the x-axis there are the US Occupy pages and on the y-axis the corresponding number of posts shared by the page administrators (blue) and by common users (green). On the total of posts, $27\%$ is made by page administrators and $73\%$ by common users. However there is low correlation $(0.37)$ between the two sharing patterns.}\label{admin2}
\end{figure}

Regardless of the page, information can be shared by its administrators, and hence appears as shared by the page itself, or by common users. Posts shared by page administrators -- i.e. admin posts -- are $27\%$ of the total (202K) against the $73\%$ (551K) from common users -- i.e., not admin posts -- and, on a page scale, there is a weak correlation $(0.37)$ between the number of posts made by admin and not admin users (Figure \ref{admin2}). 

\begin{figure}
\centering
\includegraphics[scale=0.40]{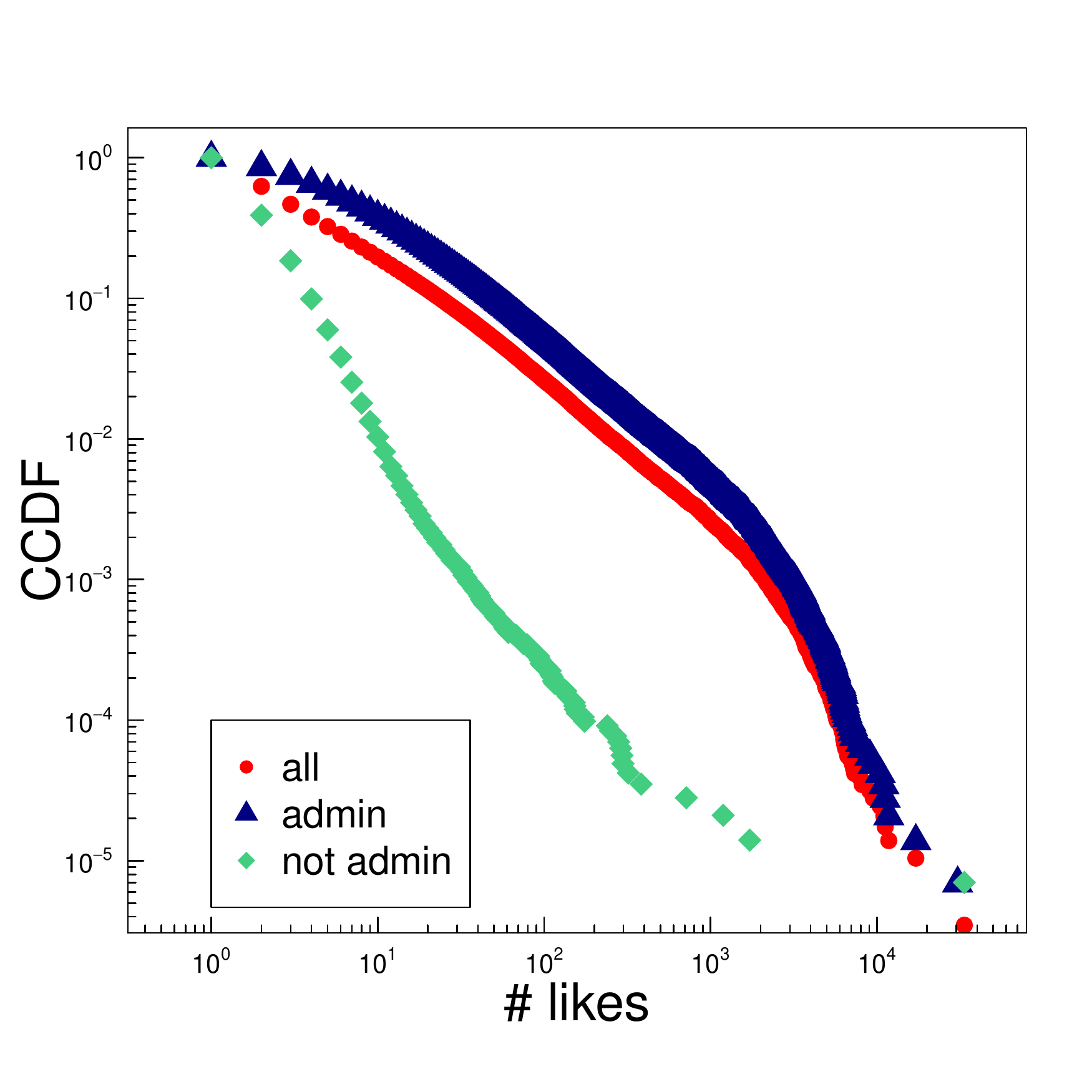}
\caption{\textbf{Likes activity on posts by admin, not admin, and all users.} The CCDF for the number of likes reached by all users posts (red), admin posts (blue), and not admin posts (green) shows a significantly bigger probability of high number of likes for the admin posts rather than not admin posts. Considering that just about one third of all the posts is shared by page administrators, we got an evident disproportion, that may be explained by the higher visibility reserved to admin posts.}\label{admin1}
\end{figure}

The number of likes to a post from the two categories is power law distributed with a quantitative difference: while the majority of posts is in the not admin category, they get less than $6\%$ of total likes, as shown in Figure \ref{admin1}. 
Number of posts by each category and number of likes reached are not proportional. This disproportion is due to the fact that, on Facebook public pages, posts by common users become soon hidden, while those from pages admin get higher visibility.
Activity on posts is power law distributed. An interesting quantitative difference between posts shared by page administrators and common users emerges.

\subsection{Users Activity}
Our dataset includes 618K active users. A user is considered to be active if she liked at least one post. 
We identified two main users categories: {\itshape occasional} -- i.e., users that made less than $5$ total likes -- and {\itshape habitual} -- i.e., users that made at least $5$ total likes.

Such a preliminary distinction, coherently with the heavy-tailed distributions of the users activities, counts $183K$ habitual users (about $30\%$ of the total). 
As a further step we define {\itshape polarized} users -- i.e.,  a habitual user is polarized on a given page if she has at least $95\%$ of her likes on that page. 
Such a thresholding classification detects about $128K$ polarized users ($21\%$ of the total), that produced $3.2M$ likes and $198K$ comments -- i.e., respectively the $62\%$ and $15\%$ of the overall likes and  comments. 

\begin{figure}
\centering
\includegraphics[scale=0.40]{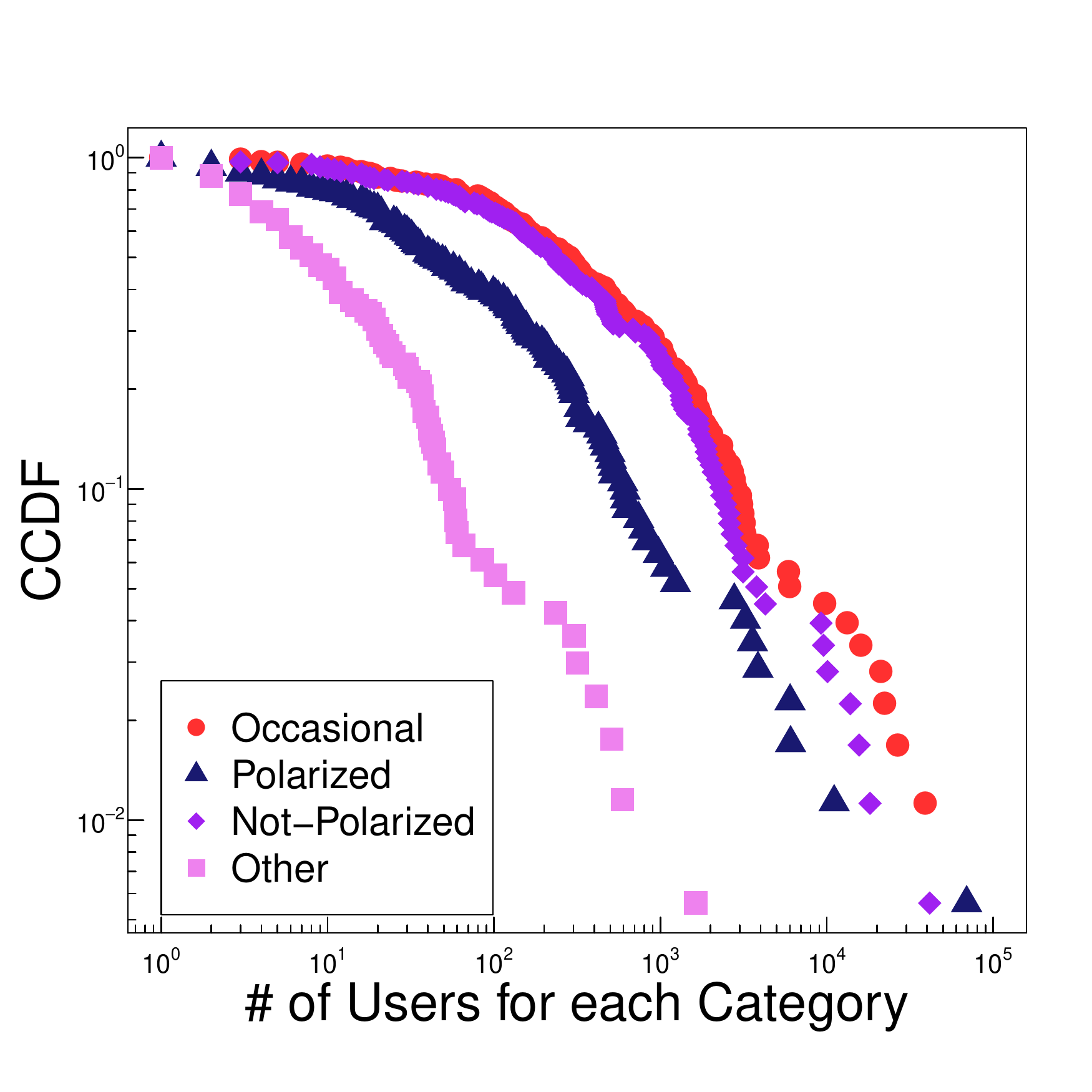}
\caption{\textbf{CCDF of different users' categories on all pages.} The CCDF of the number of active users for each categories, computed for every single page, shows how the majority of the liking activity is made by occasional (red) and not polarized (purple) users. Remember that a user is active on a page if she liked at least one post from that page. However the activity of polarized users (blue) is still significant, while users polarized on other pages (violet) represent a marginal portion of the total likes.}\label{categories}
\end{figure}

\begin{figure*}
\centering
\includegraphics[width=0.70\textwidth]{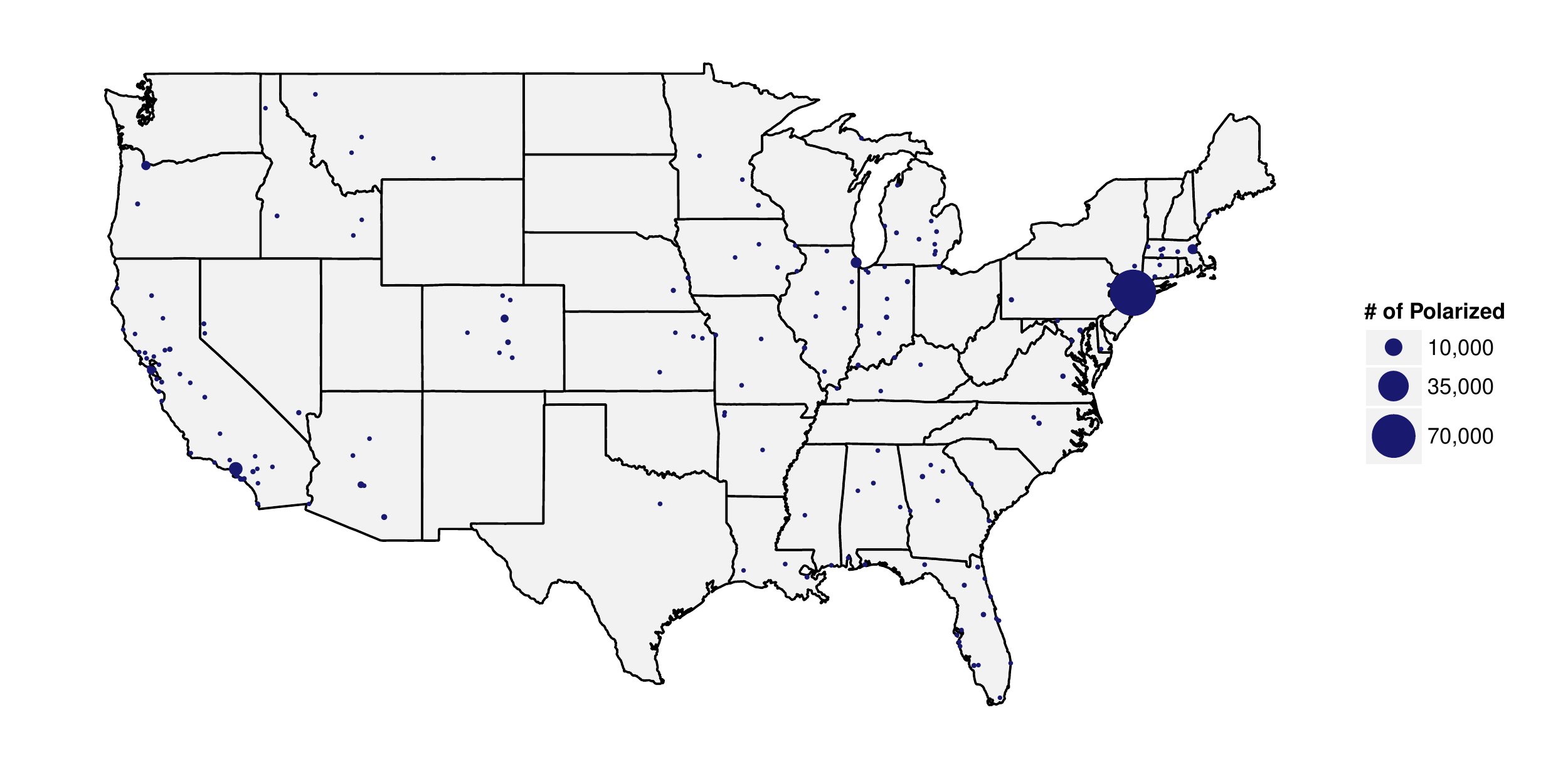}
\caption{\textbf{Geospatial distributions of the polarized users on pages.} Representation of polarized users' communities for all pages: the size of the dots is proportional to the size of the communities and emphasizes the correlation between the number of polarized users and the number of users, posts, likes, comments, and reshares for each page (see also Figure \ref{maps} and \ref{reshares}). }\label{map_polarized}
\end{figure*}

Figure \ref{categories} shows the CCDF of the number of users from all different categories on a given page -- i.e., occasional, polarized, not polarized, and polarized on a different page --, they all show a power law decay. Moreover, the size of polarized communities, for each page, is shown on the US map in Figure \ref{map_polarized}.

\begin{figure}
\centering
\includegraphics[scale=0.35]{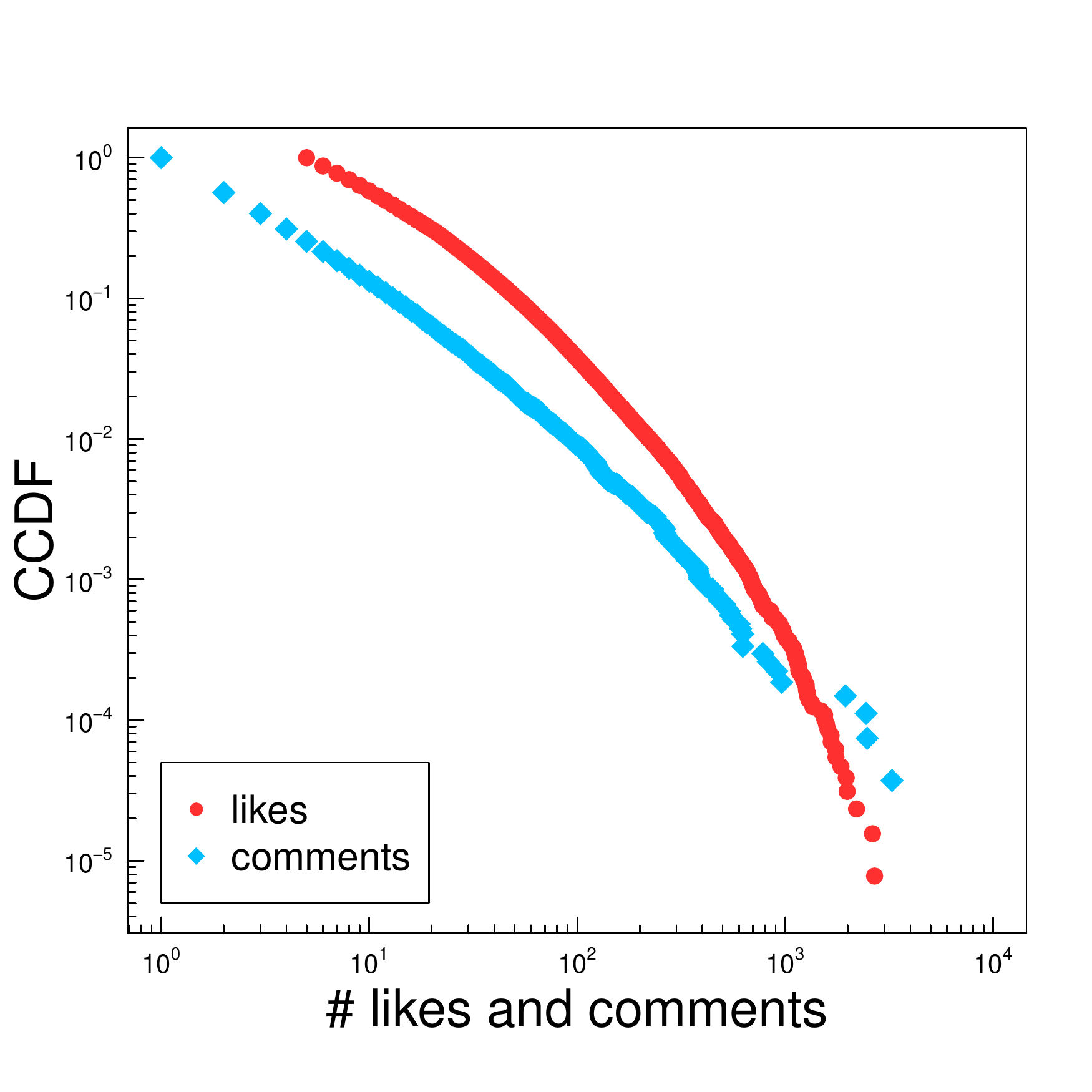}
\caption{\textbf{Activity patterns for polarized users.} The CCDF of liking (red) and commenting (sky blue) activity for polarized users shows again power law distributions. Comparing it with Figure \ref{post_activity1}, where the number of likes and comments are computed on all users, we can notice that the quantitative difference between red and sky blue curves is smaller and that, when a level of about $1000$ likes (comments) is reached, the trend starts to revert leading to a higher probability of commenting rather than liking (for polarized users).}\label{polarized_activity}
\end{figure}

Regarding the activity of polarized users we observe a power law distribution for both number of likes and comments (Figure \ref{polarized_activity}). Notice that only about $21\%$ of polarized users commented at least once.
The Pearson correlation coefficient computed on the number of polarized users, total users, posts, likes, comments, and shares, provides high correlations levels for all the combinations of measures, as shown in Table \ref{pol_tab}.

\begin{table}
\centering
{\tiny{
\begin{tabular}{l| c c c c c c}
\hline
& Polarized & Users & Posts & Likes & Comments & Shares \\ [0.5ex] 
\hline
Polarized& 1 & 0.993 & 0.948 & 0.998 & 0.978 & 0.675 \\
Users & 0.993 & 1 & 0.974 & 0.997 & 0.991 & 0.739 \\ 
Posts & 0.948  & 0.974 & 1 & 0.963 & 0.987 & 0.840 \\
Likes & 0.998 & 0.997 & 0.963 & 1 & 0.987 & 0.714 \\
Comments & 0.978 & 0.991 & 0.987 & 0.987 & 1 & 0.769 \\
Shares & 0.675 & 0.739 & 0.840 & 0.714  & 0.769 & 1 \\ 
\hline
\end{tabular}
}}
\caption{\textbf{Correlations among number of polarized users, total users, posts, likes, comments, and reshares.} Pearson correlation coefficient measures high levels of correlation within the considered elements -- i.e. number of total users, polarized users, posts, likes, comments, and reshares.}\label{pol_tab} 
\end{table}
Our findings, coherently with the heavy-tailed distributions of users activities,  show a dominant role of occasional and not polarized users in the information consumption and a differentiation of the probability of liking and commenting for polarized users depending on their level of activity.


\subsection{Backbone of Interaction Patterns}
\begin{figure}
  \centering
  \subfigure[ \textbf{$\alpha =0.01$}]
    {\includegraphics[width=0.45\textwidth]{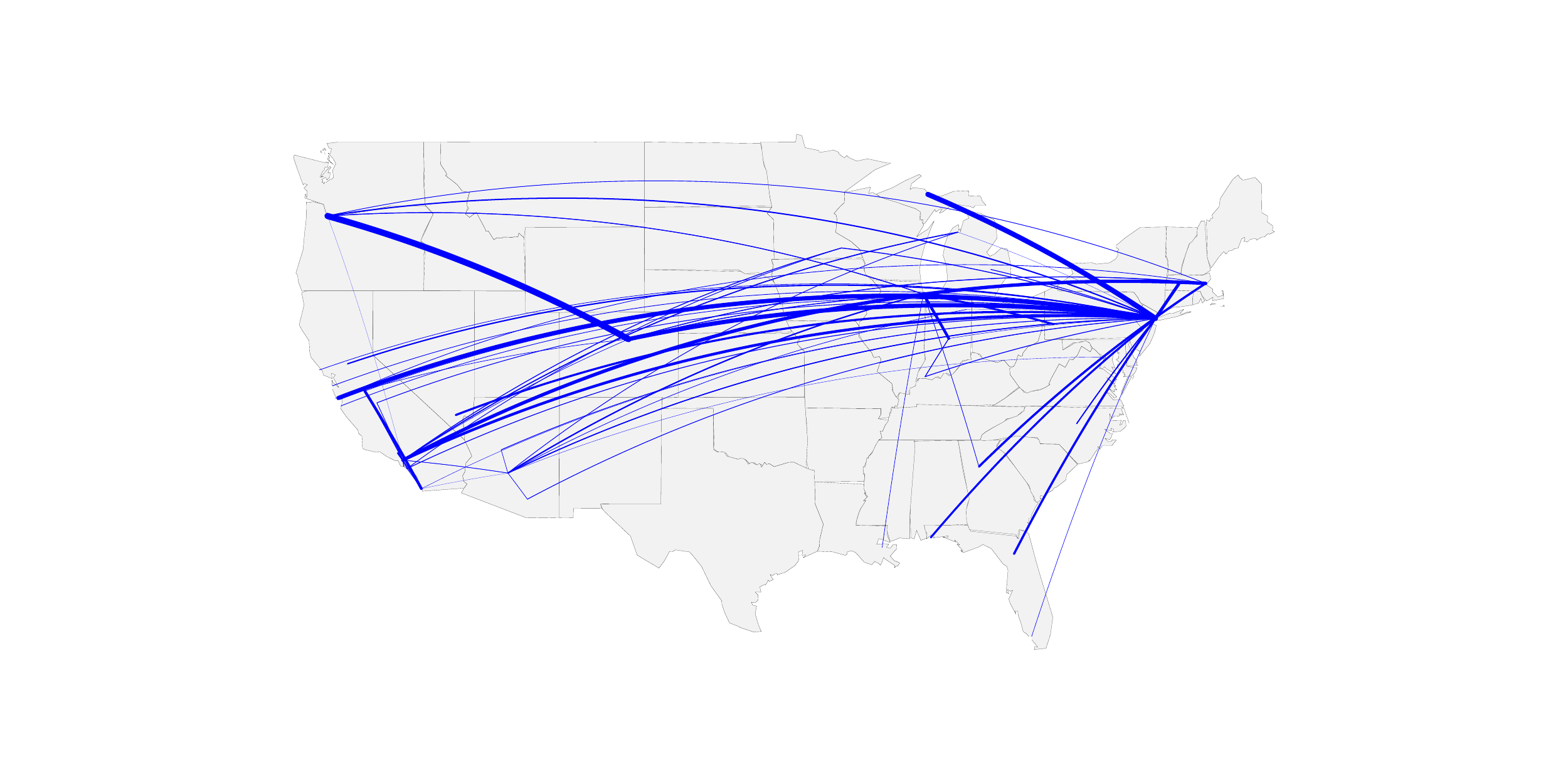}} 
  \hspace{1mm}
  \subfigure[\textbf{$\alpha =0.05$}]
   {\includegraphics[width=0.45\textwidth]{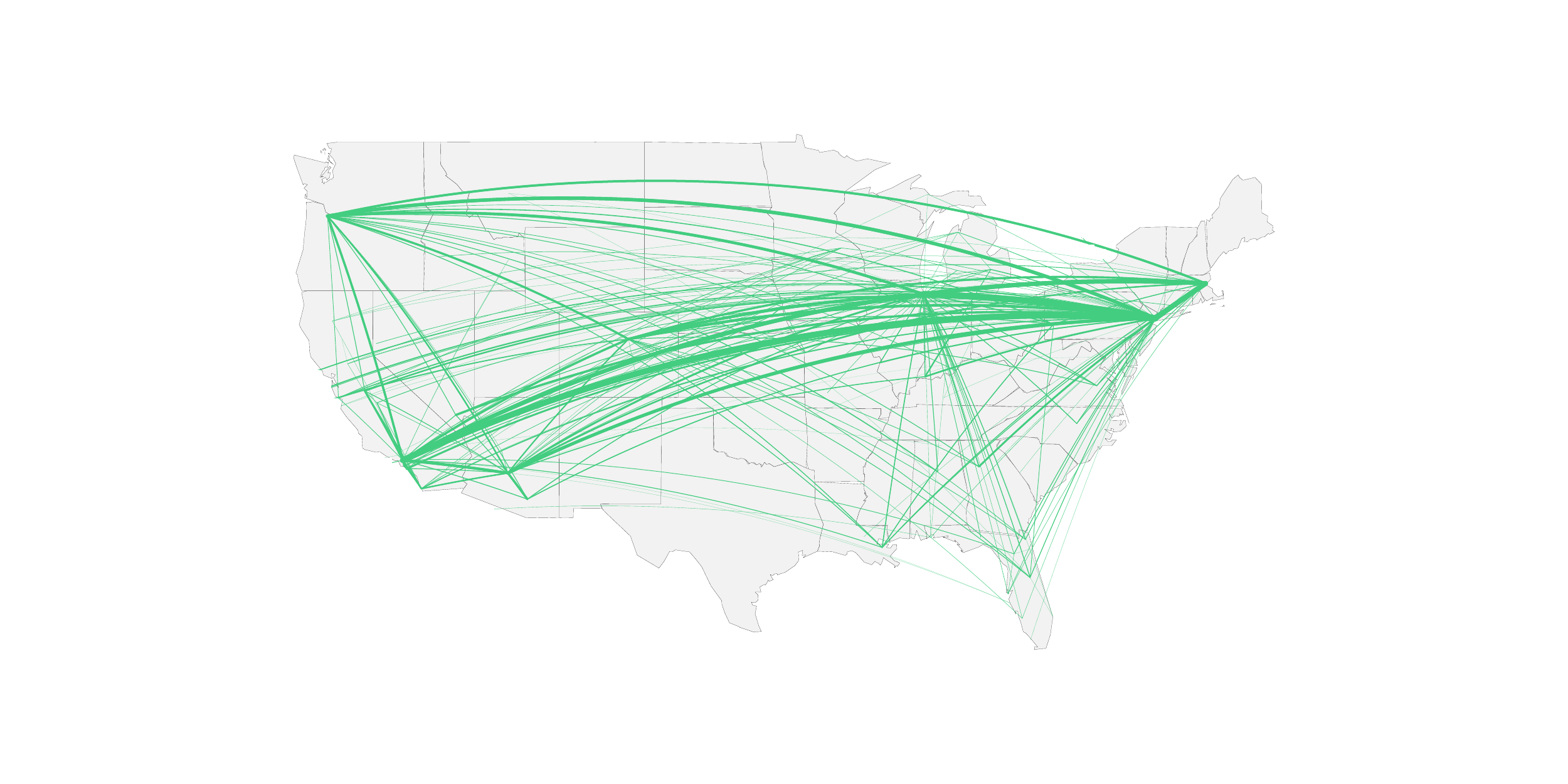}} 
\caption{\textbf{Backbone structure for the pages-reshares network}, for $\alpha=\{0.01,\,0.05\}$. The Backbone Extraction Algorithm, applied to the pages-reshares network, preserves $26.58\%$  of the total edge weight for $\alpha=0.01$ and $42.7\%$ for $\alpha=0.05$. We can notice the absence of geographical correlation in the information spreading and the emergence of US major cities -- New York, Los Angeles, Chicago, Boston, Portland, Phoenix and Denver -- as information diffusion leaders.}\label{backbone1}
 \end{figure}

\begin{figure}
  \centering
  \subfigure[ \textbf{$\alpha =0.01$}]
    {\includegraphics[width=0.45\textwidth]{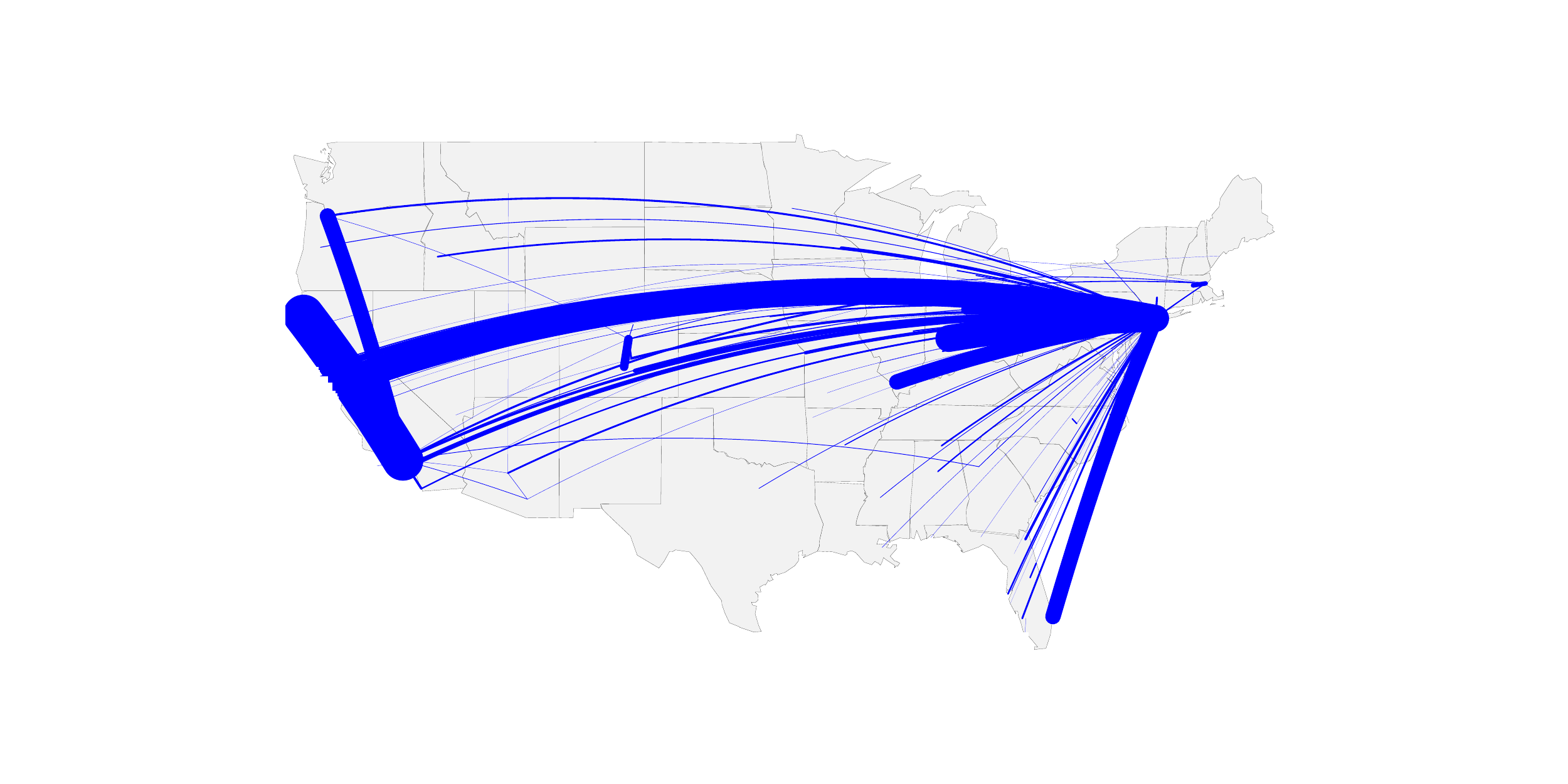}} 
  \subfigure[\textbf{$\alpha =0.05$}]
   {\includegraphics[width=0.45\textwidth]{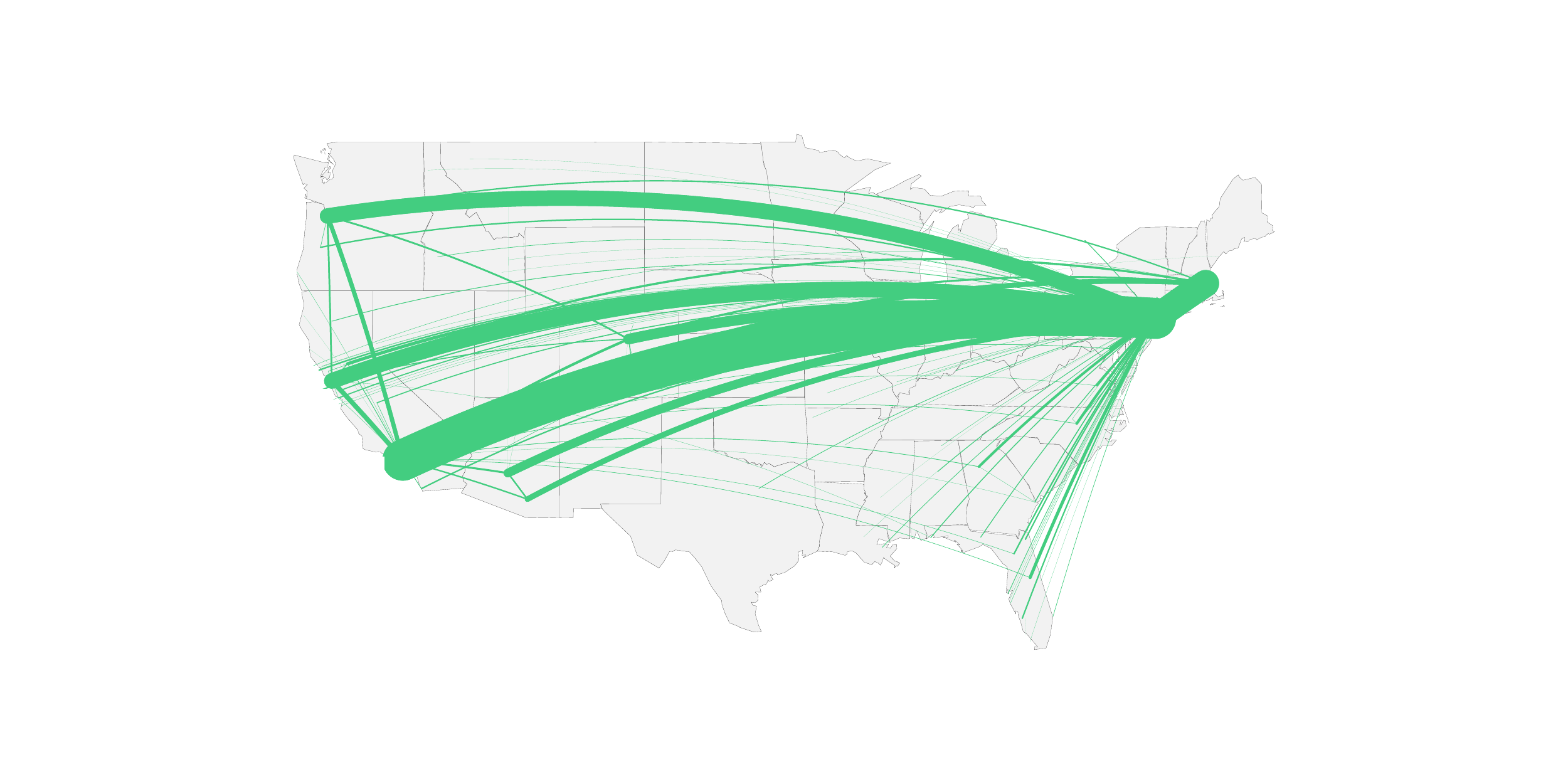}} 
      \caption{\textbf{Backbone structure for the pages-common users network}, for $\alpha=\{0.01,\,0.05\}$. The Backbone Extraction Algorithm, applied to the pages-common users network, preserves $64.95\%$ of the total edge weight for $\alpha=0.01$ and $70.65\%$ for $\alpha=0.05$. We can notice the absence of geographical correlation in the activity of polarized users across pages and the emergence of US major cities -- New York, Los Angeles, Boston, Portland and San Francisco -- as polarized users' activity exchanger.}\label{backbone2}
 \end{figure}

We want to understand if the geographical affiliation of pages or polarized users in the Occupy movement influence the diffusion of information.
Hence, we have applied a {\itshape Backbone Extraction Algorithm}~\cite{Backbone} to two different networks, as mentioned before. 
Figure \ref{backbone1} shows on the US map the results for the backbone extraction applied to the pages-reshares network, for two different levels of significance $\alpha=\{0.01,\,0.05\}$. These results provide a clear image of the absence of geographical correlation in the resharing patterns. Moreover, pages corresponding to the major US cities\footnote{For both levels of significance, the following cities emerge as information spreading leaders: New York, Los Angeles, Chicago, Boston, Portland, Phoenix and Denver.} emerge as leaders in the information spreading.
Figure \ref{backbone2} illustrates the multi-scale backbone structure for the pages-common users network, for the two levels of significance $\alpha=\{0.01,\,0.05\}$. The links correspond to the activity of users polarized on one page inside another page. 
Again, no geographic correlation is present, and five main cities showing an exchange of polarized users' activity may be identified: New York, Los Angeles, Boston, Portland and San Francisco.
 
Hubs corresponding the US major cities drive the overall activity of the movement, the diffusion of contents online and serve as coordination points for all other pages. 

\section{Conclusion}
In this paper we explore the case of online political movements, that coordinate and interact through social media even more often with the advent of the World Wide Web \cite{mccaughey2003cyberactivism}. 
Our focus is to characterize information consumption patterns by identifying different actors according to their interaction patterns with pages. As the Occupy movement online presents a geographical diversification of groups we address geographical patterns behind information diffusion.
Taking into account all the posts from $179$ US public Facebook pages about the Occupy Movement during the time span September $2011$-February $2013$, we analyze users activity on pages and posts. We find high levels of correlation for the number of users, posts, likes, comments, and reshares on each page (that are all power law distributed) and different activity patterns for posts with different post types or shared by users from different categories. Then we divide the users in two categories, habitual and occasional, according to the total number of likes they made, and we further label habitual users with at least $95\%$ of their liking activity on one particular page as polarized in that page. We find that the number of polarized users for each page is positively correlated with all the measures mentioned above and that the most of the liking activity is performed by occasional or not polarized users. Moreover, we extract the multi-scale backbone for two networks, the pages-reshares network and the pages-common users network, in order to analyze geographical patterns in the information diffusion and polarized users activity.
Our analysis reveals that activities online are not locally coordinated by geographically close pages. Indeed, pages linked to major US cities -- e.g., New York, Los Angeles, Chicago, Boston, San Francisco -- drive the diffusion of contents online and serve as coordination points for all other pages, which perform a minor activity in the system.

\section{Acknowledgments}
Funding for this work was provided by EU FET project MULTIPLEX nr. 317532 and SIMPOL nr. 610704. The funders had no role in study design, data collection and analysis, decision to publish, or preparation of the manuscript.

	
\bibliographystyle{unsrt}
\bibliography{occupy}

\end{document}